\documentclass[sigconf,nonacm]{acmart}

\usepackage{booktabs}
\usepackage{amsmath}
\usepackage{graphicx}
\usepackage{microtype}
\usepackage{listings}
\usepackage[most]{tcolorbox}

\newtcblisting{promptbox}{
  breakable,
  enhanced,
  listing only,
  colback=black!2.5,
  colframe=black!18,
  boxrule=0.35pt,
  arc=2pt,
  borderline west={1.8pt}{0pt}{black!65},
  left=7pt,
  right=5pt,
  top=4pt,
  bottom=4pt,
  before skip=0.45em,
  after skip=0.7em,
  listing options={
    basicstyle=\ttfamily\footnotesize,
    breaklines=true,
    breakatwhitespace=false,
    columns=fullflexible,
    keepspaces=true,
    showstringspaces=false,
    aboveskip=0pt,
    belowskip=0pt
  }
}

\renewcommand\footnotetextcopyrightpermission[1]{}

\title{DM-Lab at TREC 2026 RAG: Multi-Query Retrieval, Ensemble Judging, and Claim-Pack Generation}

\author{Elyas Irankhah}
\authornote{Co-corresponding authors.}
\email{elyas.irankhah@yale.edu}
\affiliation{%
  \department{Department of Emergency Medicine}
  \institution{Yale University School of Medicine}
  \city{New Haven}
  \state{CT}
  \country{USA}
}

\author{Muhammad Arif}
\authornotemark[1]
\email{muhammad.arif@yale.edu}
\affiliation{%
  \department{Department of Emergency Medicine}
  \institution{Yale University School of Medicine}
  \city{New Haven}
  \state{CT}
  \country{USA}
}

\author{Sreeraj Ramachandran}
\email{sreeraj.ramachandran@yale.edu}
\affiliation{%
  \department{Department of Emergency Medicine}
  \institution{Yale University School of Medicine}
  \city{New Haven}
  \state{CT}
  \country{USA}
}

\author{Samah Fodeh-Jarad}
\email{samah.fodeh@yale.edu}
\affiliation{%
  \department{Department of Emergency Medicine}
  \institution{Yale University School of Medicine}
  \city{New Haven}
  \state{CT}
  \country{USA}
}

\begin{abstract}
We present DM-Lab-RAG-2026, a framework developed for the TREC 2026 Retrieval-Augmented Generation (RAG) track. The framework integrates large-scale evidence retrieval and citation-grounded answer generation over the 553-million-document Climb\-Mix-400B corpus. The retrieval architecture combines multi-query BM25, reciprocal rank fusion, ensemble relevance assessment, listwise reranking, and adaptive retrieval depth. The generation pipeline employs a claim-first strategy to extract evidence-linked atomic claims, construct cited responses, and verify source support. Across 22 development topics, the retrieval system achieved a mean nDCG@30 of 0.604. In the official evaluation, it attained a set precision of 0.455 and a manual nugget gain of 0.910. Additional experiments examine the influence of reranking, evidence selection, and citation verification on system performance. The evaluation demonstrates the potential of diversified lexical retrieval and language-model-based relevance assessment for large-scale evidence discovery. It further identifies evidence preservation and structured answer construction as key considerations for reliable RAG. This work provides a reproducible framework for investigating retrieval effectiveness and evidence-grounded generation.
\end{abstract}

\ccsdesc[500]{Information systems~Retrieval models and ranking}
\ccsdesc[300]{Information systems~Question answering}

\keywords{retrieval-augmented generation, TREC, reciprocal rank fusion, BM25 }

\begin{document}

\maketitle


\section{Introduction}

Retrieval-augmented generation (RAG) enables language models to produce knowledge-intensive answers using externally retrieved evidence~\cite{lewis2020rag}. Its effectiveness depends on identifying relevant documents, covering essential information, and generating coherent responses with reliable citations~\cite{gao2023alce,liu2024lostmiddle}. These requirements become particularly challenging for complex narrative queries involving multiple information needs. Effective retrieval alone cannot guarantee that relevant evidence is preserved, accurately synthesized, and appropriately attributed during answer generation.

The TREC 2026 RAG track provides a large-scale setting for investigating these challenges using ClimbMix-400B, a corpus containing approximately 553 million documents~\cite{diao2025nemotron,sharifymoghaddam2026climbmix}. The track evaluates evidence retrieval and end-to-end answer generation, emphasizing factual coverage and citation support~\cite{trec2026guidelines}. Existing approaches improve individual stages through query expansion, rank fusion, language-model reranking, and relevance assessment~\cite{gao2023hyde,cormack2009rrf,sun2023chatgpt,upadhyay2024umbrela}. Integrating these techniques into an effective RAG pipeline requires careful coordination between evidence selection and answer construction.

\begin{figure*}[t]
  \centering
  \includegraphics[width=\textwidth]{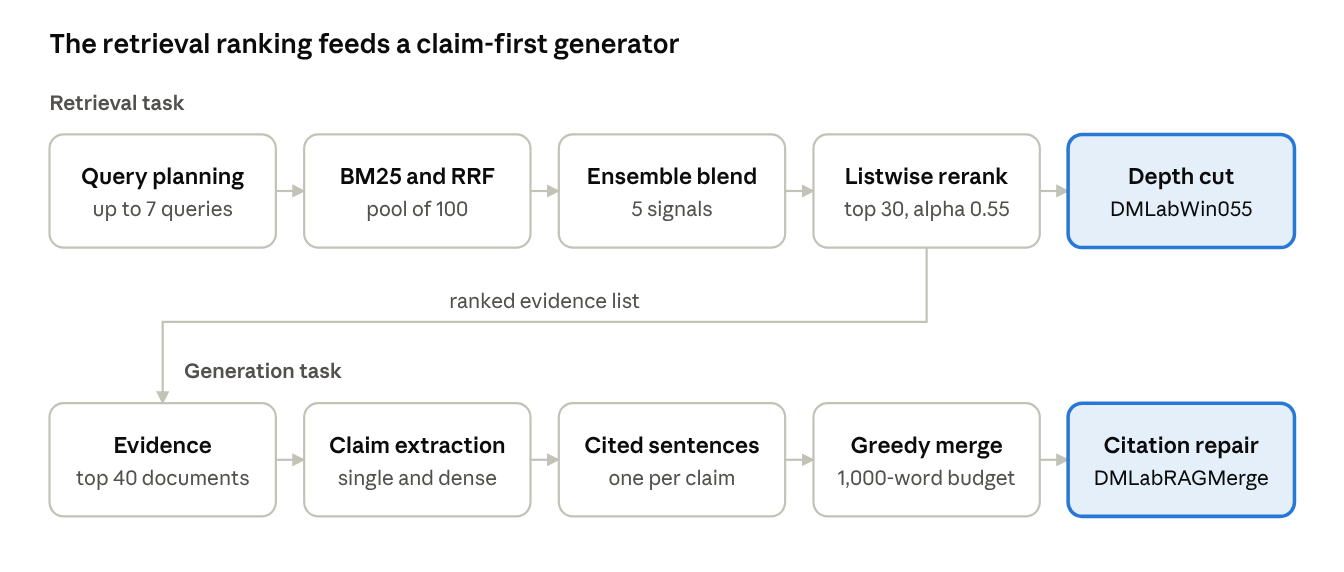}
  \caption{End-to-end submitted pipeline. Multiple lexical query views feed the retrieval run \texttt{DMLabWin055}. The top 40 documents before depth truncation feed the claim-first generation run \texttt{DMLabRAGMerge}.}
  \Description{Two-row pipeline diagram. The retrieval row proceeds from query planning through BM25 and reciprocal rank fusion, ensemble blending, listwise reranking, and a depth cut. The generation row consumes the top 40 evidence documents, extracts claims, creates cited sentences, merges them under a word budget, and repairs citations.}
  \label{fig:pipeline}
\end{figure*}

To address these challenges, we develop two fully automatic systems for the TREC 2026 RAG track. The retrieval system, \texttt{DMLabWin\allowbreak055}, combines multi-query BM25, reciprocal rank fusion, ensemble relevance assessment, listwise reranking, and adaptive retrieval depth. The generation system, \texttt{DMLabRAG\allowbreak Merge}, performs evidence-linked claim extraction, citation-grounded answer construction, and source-support verification. Figure~\ref{fig:pipeline} illustrates the proposed architecture. We assess both pipelines through component-level experiments and official TREC evaluations to examine the relationship between evidence selection and answer quality. The source code and experimental resources are available on \url{https://github.com/Elyasirankhah/TREC-RAG-2026}.

\subsection{Background and Related Work}

Research on retrieval-augmented generation has increasingly focused on reproducible evaluation and the relationship between retrieval effectiveness and answer quality. The Ragnar\"ok framework introduced reusable baselines and standardized interfaces for evaluating RAG systems~\cite{pradeep2025ragnarok}. BERGEN demonstrated the value of controlled comparisons across retrievers, rerankers, and language models~\cite{rau2024bergen}. The TREC 2025 RAG track expanded evaluation through retrieval, generation, and relevance-assessment tasks~\cite{trec2025overview}. Prior research establishes a methodological foundation for evaluating retrieval quality and examining how language models utilize retrieved evidence~\cite{es2024ragas,chen2024rgb}. RAGFlip examines query-level performance regressions in retriever upgrades, highlighting reliability considerations that aggregate retrieval metrics may obscure~\cite{irankhah2026ragflip}.

Advances in evidence acquisition have incorporated language-model-assisted query reformulation and relevance assessment. Query\-2doc demonstrated that generated pseudo-documents can improve lexical retrieval without additional retriever training~\cite{wang2023query2doc}. Large-scale studies of LLM-based relevance judgments have reported strong correlations with human-derived system rankings~\cite{upadhyay2024largescale}. Such approaches support diversified lexical retrieval and automated relevance assessment. Their application also raises questions about judgment consistency, relevance thresholds, and the selection of evidence across multiple information needs.

Answer evaluation extends beyond document relevance to factual completeness and source attribution. Nugget-based methods assess the coverage of essential facts, while citation-support evaluations measure whether attributed documents substantiate generated statements~\cite{pradeep2025nugget,thakur2025support}. VeriScore evaluates the factual accuracy of verifiable claims across long-form generation tasks~\cite{song2024veriscore}. ARES and RAGChecker introduce fine-grained evaluation methods for assessing component quality and identifying failures across retrieval and generation stages~\cite{saadfalcon2024ares,ru2024ragchecker}. Building on prior research on deterministic evidence grounding and multi-pass alignment~\cite{irankhah2026archehr}, our work investigates how evidence is retained during document selection, claim extraction, sentence assembly, and citation filtering. The analysis focuses on the relationship between retrieved evidence, factual coverage, and citation support in long-form answer generation.

\section{Task analysis}

The 2026 guidelines reward three outcomes: pairwise answer quality, nugget coverage, and citation support~\cite{trec2026guidelines,thakur2025support}.
Each RAG system retrieves its own evidence because the augmented-generation-only task was removed.
Table~\ref{tab:rules} maps the main rules to the resulting design choices.

\noindent\begin{minipage}{\columnwidth}
  \captionof{table}{Design consequences of the task rules.}
  \label{tab:rules}
  \small
  \setlength{\tabcolsep}{4pt}
  \begin{tabular}{@{}p{0.46\columnwidth}p{0.46\columnwidth}@{}}
    \toprule
    Rule & Consequence \\
    \midrule
    119 narratives; IDs and text preserved exactly &
    Copy metadata from the official file and check it automatically. \\
    \midrule
    Retrieval depth $k$ chosen per narrative; submit all and only useful documents &
    Depth is itself a prediction, so the stopping rule matters. \\
    \midrule
    Answers limited to 1,024 words by \texttt{str.split()} &
    Budget selection becomes a coverage problem. \\
    \midrule
    Answer objects carry 0 to 3 citations, strongest first &
    Each sentence needs its best evidence named first. \\
    \midrule
    An uncited object scores 0 for citation recall &
    Removing a bad citation can raise precision and leave recall unchanged, or zero an object that loses its last citation. \\
    \midrule
    Uncited references carry no penalty &
    The reference list may be kept or pruned. \\
    \midrule
    Pairwise battles plus nugget scoring~\cite{pradeep2025nugget} &
    Answers need both coverage and readable synthesis. \\
    \bottomrule
  \end{tabular}
\end{minipage}

\vspace{0.75em}

Two features of this evaluation drove the analysis.
Nuggets reward factual density.
A pairwise assessor may also reward organization and direct advice.
A strategy tuned for one may not win the other.
Weighted citation precision and recall also respond differently to citation pruning.
During development we approximated support with $w(s,d)\in\{0,0.5,1\}$ for no, partial, or full support of sentence $s$ by document $d$.
If $C$ is the set of citation pairs, $S$ the set of answer objects, and $D(s)$ the documents cited by $s$, with an empty $D(s)$ scoring 0, then
\begin{align}
\mathrm{Prec} &= \frac{1}{|C|}\sum_{(s,d)\in C} w(s,d), \\
\mathrm{Rec} &= \frac{1}{|S|}\sum_{s\in S}
  \frac{1}{\max(1,|D(s)|)}\sum_{d\in D(s)} w(s,d).
\end{align}

Test narratives range from 13 to 206 words, with a mean of 83.1.
The median narrative contains two question marks.
Development data contains 22 RAG 2025 narratives~\cite{trec2025overview}.
It includes nuggets and three projected UMBRELA qrels.
We also used 30 ResearchRubrics prompts.
Those qrels are model-generated over pooled candidates, not the official 2026 judgments~\cite{trec2026guidelines}.

\section{Retrieval system: DMLabWin055}

\texttt{DMLabWin055} retrieves documents with BM25 from the shared \texttt{climbmix-400b} index, accessed through the Pyserini REST service.
It builds no index of its own.
The retrieved documents are scored by WIN, the ensemble weights selected on the development topics (Table~\ref{tab:weights}).
A listwise reranker then reorders the top 30 documents, and the submitted score gives this order weight $\alpha=0.55$.

\subsection{Query planning}

\texttt{gpt-4.1-mini} turns each narrative into as many as seven BM25 queries.
The query set contains the original narrative and a bag of up to 24 entities and keywords.
It also contains one concatenated query of at most 120 words, built from subquestions and keywords.
The remaining two to four queries are hypothetical evidence passages of 40--80 words, in the style of HyDE~\cite{gao2023hyde}.
Some code labels refer to SPLADE and dense retrieval, but those branches also run BM25 over rewritten queries.
We therefore describe them as lexical query views.

\subsection{Candidate fusion}

Each query retrieves 200 documents.
We remove near-duplicates before fusion.
Two documents are duplicates when their first 200 characters match or when token Jaccard reaches 0.9 over their first 1,200 characters.
This filter is consequential because a published analysis estimates that 39.6\% of ClimbMix documents have a duplicate~\cite{sharifymoghaddam2026climbmix}.
Unweighted reciprocal rank fusion with $k=60$~\cite{cormack2009rrf} then forms a pool of 100 documents.
For query views $Q$, with rank $r_q(d)$ of document $d$ in view $q$,
\begin{equation}
\mathrm{RRF}(d)=\sum_{q\in Q}\frac{1}{60+r_q(d)}.
\end{equation}

\subsection{Sub-narrative generation}

The UMBRELA prompt~\cite{upadhyay2024umbrela} requires atomic sub-narratives. An LLM generates two independent decompositions, which are subsequently merged. Few-shot examples are drawn from the official TREC 2025 RAG narratives~\cite{trec2025overview}. Each decomposition contains six to ten atomic facets.

\subsection{Ensemble score}

Signals are normalized to $[0,1]$ and combined with the frozen development weights in Table~\ref{tab:weights}.
The weights came from a grid search over mean nDCG@30 across the three development qrels, with a leave-one-topic-out check.
The repository records mean nDCG@30 of 0.5911 under leave-one-topic-out evaluation.
The corresponding in-sample value is 0.5992 for the dual-judge blend before listwise reranking.
The test recipe uses the same weights and redistributes the weight of any missing signal.
The published test command omits the development coverage cache.
The coverage weight is therefore redistributed over retrieval rank, the pointwise grade, and the two UMBRELA judges.

\begin{table}[t]
  \caption{Frozen development ensemble. On test, a missing signal's weight is redistributed.}
  \label{tab:weights}
  \begin{tabular}{cll}
    \toprule
    Signal & Source & Weight \\
    \midrule
    $r$  & Rank percentile after fusion & 0.3 \\
    $g$  & \texttt{gpt-4.1-mini} grade, 0 to 100 & 0.1 \\
    $u_q$ & UMBRELA grade, Qwen3.5-9B & 0.4 \\
    $u_m$ & UMBRELA grade, Ministral-3-14B & 0.1 \\
    cov & Coverage pseudo-UMBRELA & 0.1 \\
    \bottomrule
  \end{tabular}
\end{table}

\subsection{Listwise rerank}

\texttt{gpt-5.6-sol} reorders the top 30 documents.
The model sees up to 1,800 characters from each document and the generated sub-narratives.
The prompt favors passages that answer more facets in detail~\cite{sun2023chatgpt}.
For a document in that head, the submitted score is
\begin{equation}
s(d)=0.55\cdot \mathrm{pct}_{\mathrm{list}}(d)+0.45\cdot \mathrm{pct}_{\mathrm{ens}}(d).
\end{equation}
Documents below the head keep their ensemble order.

\subsection{Variable depth}

The submitted run keeps a document when the maximum Qwen or Ministral grade is at least 2.
The stopping policy has a floor of one document and a ceiling of 100.
Table~\ref{tab:depth} reports the file that was uploaded.
Mean depth is 56.0 documents.
One narrative was submitted with a single document.
An earlier development policy used a floor of 10, the uploaded test file uses a floor of 1.

\begin{table}[t]
  \caption{Submitted retrieval depth.}
  \label{tab:depth}
  \small
  \begin{tabular}{lrrrrr}
    \toprule
    Run & Topics & Min & Median & Mean & Max \\
    \midrule
    \texttt{DMLabWin055} & 119 & 1 & 59 & 56.0 & 99 \\
    \bottomrule
  \end{tabular}
\end{table}

\section{Generation system: DMLabRAGMerge}

The generator separates evidence harvesting from answer writing.
Each emitted sentence begins as a claim tied to specific documents.
Evidence is the top 40 documents of the listwise run, before the variable-depth cut.
These documents are split into sentences of 50--420 characters.
We score each sentence by its overlap with the narrative and sub-narratives.
A round-robin pass across documents retains up to 80 evidence sentences.

\texttt{gpt-5.6-sol} performs the first atomic-claim extraction pass.
Every claim must cite an evidence identifier.
\texttt{gpt-4.1-mini} performs a second, denser pass.
It adds overlapping windows of 18 evidence sentences and targeted extraction for sub-narratives whose terms overlap the current claims by less than 0.35.
It keeps up to 90 claims and removes near-duplicates at 0.85 content-term overlap.
Each surviving claim becomes a sentence of 12--28 words.
A sentence may use up to three citations drawn only from its source claim.

An unsubmitted anchor variant reserved 750 words for the dense pass.
At each step a candidate sentence $s$ is scored against the draft's content terms $T$ and the term set $A$ of each sub-narrative,
\begin{equation}
\mathrm{score}(s)=|t(s)\setminus T|+3\sum_{A}\max\Bigl(0,\tfrac{|A\cap(T\cup t(s))|-|A\cap T|}{|A|}\Bigr).
\end{equation}

The highest-scoring sentence that fits is added until 1,000 words are used, or no candidate scores above zero.

We use \texttt{gpt-4.1-mini} for citation repair.
It assigns each pair one of three labels: support, partial support, or no support.
The model sees up to 4,000 characters from the cited document.
Citations labeled no support are removed.
Partial-support citations are kept.
Table~\ref{tab:gen} summarizes the uploaded file.
Of 8,003 answer objects, 1,948 have an empty citation list.

\begin{table}[t]
  \caption{Submitted answers.}
  \label{tab:gen}
  \small
  \begin{tabular}{lrrrr}
    \toprule
    Run & Words & Objects & Refs & Uncited \\
    \midrule
    \texttt{DMLabRAGMerge} & 996 & 67.3 & 32.1 & 16.4 \\
    \bottomrule
  \end{tabular}
\end{table}

Merged sentences appear in selection order.
This policy can produce a list of facts with weak transitions and limited synthesis.
Repair can leave uncited objects.
Each such object receives zero weighted citation recall.
Citation order follows extraction order, although the guidelines request strongest support first.

\section{Development Results}

We evaluated the retrieval system on 22 TREC 2025 development topics using three relevance-judgment sets. Table~\ref{tab:ndcg} presents the ensemble and listwise reranking results. The final configuration achieved a mean nDCG@10 of 0.656 and nDCG@30 of 0.604. Both configurations retained 100 documents per topic which allow direct assessment of ranking effectiveness without variable-depth effects.

\begin{table}[t]
  \caption{Mean nDCG across three relevance judges on 22 development topics.}
  \label{tab:ndcg}
  \small
  \begin{tabular}{lrrrr}
    \toprule
    Judge & Ens.@10 & +List@10 & Ens.@30 & +List@30 \\
    \midrule
    GPT-5.5 & 0.574 & 0.627 & 0.533 & 0.550 \\
    Ministral-3-14B & 0.712 & 0.722 & 0.655 & 0.657 \\
    Qwen3.5-9B & 0.635 & 0.620 & 0.610 & 0.604 \\
    \midrule
    DM-Lab mean & 0.640 & 0.656 & 0.599 & 0.604 \\
    \bottomrule
  \end{tabular}
\end{table}

A paired bootstrap analysis with 10,000 samples assessed the contribution of listwise reranking. Mean improvements reached $+0.016$ for nDCG@10 and $+0.005$ for nDCG@30. Their respective 95\% confidence intervals were $[-0.006, 0.037]$ and $[-0.003, 0.012]$. Reranking improved 14 of 22 topics, with the largest judge-specific gain observed under GPT-5.5. Topic-level nDCG@30 changes ranged from $+0.033$ to $-0.031$. Final per-topic scores ranged from 0.310 to 0.857, with a median of 0.634. The interpolation weight of 0.55 was selected on the same development set, so the estimated improvements require independent validation.

Additional diagnostics examined judgment coverage and consistency. Relevance assessments covered 90.2\% of the top 30 documents, with unjudged documents treated as non-relevant. The final mean nDCG@30 had a 95\% bootstrap interval of $[0.539, 0.667]$. Across 26,341 judgment pairs, exact agreement ranged from 0.382 to 0.565, and quadratic-weighted kappa ranged from 0.602 to 0.736. Ministral assigned relevance grades of at least 2 to 75.9\% of pooled documents, compared with 49.4\% for GPT-5.5. This variation illustrates the importance of judge calibration in evidence selection.

Generation experiments assessed factual coverage and citation reliability. The merged claim configuration achieved lexical vital recall of 0.549 and strict vital support of 0.363. Independent RAGDoll ~\cite{ragdoll2026}  evaluation yielded a micro strict vital score of 0.333. Citation verification increased locally assessed precision from 0.725 to 0.926, while recall remained at 0.731. An oracle using gold nuggets achieved strict vital support of 0.481, providing a reference for evidence-selection analysis. Additional quote-first extraction experiments did not improve the development criterion over the selected configuration.

\section{Official Results}

Both DM-Lab systems passed the official TREC 2026 validation. The retrieval evaluation used relevance judgments and nugget-based metrics. Documents with UMBRELA grades of at least 2 were considered relevant. Nugget gain weighted vital and okay facts by 1 and 0.5, respectively, and required complete evidence support.

\begin{table}[t]
  \caption{Official retrieval evaluation of \texttt{DMLabWin055}.}
  \label{tab:official-r}
  \small
  \begin{tabular}{lr}
    \toprule
    Metric & DM-Lab \\
    \midrule
    Set precision (82) & 0.455 \\
    Set recall (82) & 0.118 \\
    Documents (82) & 57.5 \\
    Manual nugget gain (25) & 0.910 \\
    Automatic nugget gain (25) & 0.687 \\
    Nugget effort (25) & 62.8 \\
    \bottomrule
  \end{tabular}
\end{table}

As shown in Table~\ref{tab:official-r}, \texttt{DMLabWin055} attained a set precision of 0.455 and a set recall of 0.118, with 57.5 documents submitted. Manual nugget gain reached 0.910, and automatic nugget gain reached 0.687, with a nugget effort of 62.8. The evaluation demonstrates effective evidence selection.

The generation system, \texttt{DMLabRAGMerge}, was evaluated using human pairwise preferences, rubric assessments, and nugget coverage. Table~\ref{tab:official-rag} summarizes the two NIST assessment batches, covering 17 and 25 narratives, respectively.

\begin{table}[t]
  \caption{Official generation evaluation of \texttt{DMLabRAGMerge}.}
  \label{tab:official-rag}
  \small
  \begin{tabular}{lr}
    \toprule
    Metric & DM-Lab \\
    \midrule
    BT rating, batch 1 & 975 \\
    BT rating, batch 2 & 940 \\
    Rubric vital, batch 1 & 0.536 \\
    Rubric all, batch 1 & 0.593 \\
    Manual vital, batch 2 & 0.461 \\
    Manual all, batch 2 & 0.481 \\
    Automatic vital, batch 2 & 0.077 \\
    Automatic all, batch 2 & 0.062 \\
    \bottomrule
  \end{tabular}
\end{table}

The generation evaluation showed stronger coverage under human nugget assessment than under automatic assignment. Manual vital coverage reached 0.461, while automatic vital coverage was 0.077. Human preference ratings were 975 in batch 1 and 940 in batch 2. The differences across evaluation methods highlight the challenges of preserving factual coverage and answer coherence during evidence-grounded generation.

\section{Discussion}

\subsection{What Worked}

The official evaluation supports the effectiveness of diversified lexical retrieval and LLM-based relevance assessment. The retrieval system attained a set precision of 0.455, a set recall of 0.118, a manual nugget gain of 0.910, and an automatic nugget gain of 0.687, with 57.5 documents submitted. The performance highlights the value of multi-query BM25, rank fusion, and ensemble relevance judgments for large-scale evidence discovery. The approach also eliminates the need for a custom dense index over ClimbMix, reducing indexing requirements and supporting deployment in resource-constrained environments.

\subsection{What Failed}

The generation pipeline encountered challenges in converting retrieved evidence into coherent and comprehensive answers. Claim extraction emphasized factual atomicity and source attribution, while greedy sentence selection prioritized lexical novelty. The absence of explicit discourse planning limited information organization across narrative facets. Citation filtering introduced an additional tradeoff. Removing unsupported references increased locally assessed citation precision but left 24.3\% of answer objects uncited. The outcome underscores the importance of integrating claim selection, answer organization, and citation verification.

\subsection{Implications}

The evaluation indicates that retrieval effectiveness and generation quality require distinct optimization strategies. Organizing extracted claims by narrative facet could improve answer coherence and factual coverage. Citation verification could incorporate sentence revision to preserve source attribution. The limited gains from listwise reranking also suggest opportunities to allocate computational resources toward structured answer planning and evidence verification.

\section{Limitations}

Development experiments used 22 topics for both parameter selection and evaluation, potentially introducing optimistic estimates. Model-generated relevance judgments may introduce evaluator bias. BM25 retrieval may overlook relevant documents with limited lexical overlap. Generation evaluation covered selected runs across two assessment batches. Citation verification also relied on a single LLM judge, making support assessments sensitive to model calibration.


\section{Conclusion}

We presented two fully automatic systems for the TREC 2026 RAG track, integrating large-scale lexical retrieval with claim-first answer generation. The retrieval pipeline combined multi-query BM25, rank fusion, ensemble relevance assessment, and adaptive depth selection. Official retrieval results included a set precision of 0.455 and a manual nugget gain of 0.910, with 57.5 documents submitted. The generation evaluation identified opportunities to improve evidence preservation, answer coherence, and citation support. Future work will focus on facet-aware answer planning and support-aware claim revision. Implementation details, experimental configurations, and additional analyses are provided in Appendix~\ref{app:supplement}.

\begin{acks}
We thank the TREC RAG organizers and NIST.
\end{acks}

\bibliographystyle{ACM-Reference-Format}
\bibliography{references}

\appendix

\section{Appendix}
\label{app:supplement}

This appendix provides the reproducibility details, submitted configurations, additional diagnostics, and prompt templates. The implementation is available at \url{https://github.com/Elyasirankhah/TREC-RAG-2026}.

\subsection{Submitted Configurations}

Table~\ref{tab:repro} summarizes the frozen parameters of both submitted systems.

\begin{table*}[t]
  \caption{Configuration parameters of the submitted retrieval and generation systems.}
  \label{tab:repro}
  \small
  \begin{tabular}{p{0.20\textwidth}p{0.35\textwidth}p{0.35\textwidth}}
    \toprule
    Component & Retrieval configuration & Generation configuration \\
    \midrule
    Evidence selection &
    Up to 7 BM25 queries; 200 documents/query; RRF $k=60$; pool of 100 &
    Top 40 reranked documents before depth truncation \\
    Language models &
    \texttt{gpt-4.1-mini}, Qwen3.5-9B, Ministral-3-14B &
    \texttt{gpt-5.6-sol}, \texttt{gpt-4.1-mini} \\
    Listwise reranking &
    Top 30; 1,800 characters/document; $\alpha=0.55$ &
    --- \\
    Selection constraints &
    $\max(u_q,u_m)\geq2$; depth 1--100 &
    Maximum 1,000 words \\
    Deduplication &
    Prefix matching or Jaccard $\geq0.90$ &
    Claim-term overlap $\geq0.85$ \\
    Citation verification &
    --- &
    Up to 3 citations; remove unsupported citations; retain partial support \\
    \bottomrule
  \end{tabular}
\end{table*}

\subsection{Additional Diagnostics}

\subsubsection{Generation Loss Analysis}

Development diagnostics examined evidence retention across generation stages. Vital coverage reached 0.984 at the document level and 0.710 after span selection. The submitted claim-merging system achieved strict vital support of 0.363. A gold-nugget oracle reached 0.481 but was excluded from submission. On an eight-topic evaluation, the selected configuration scored 0.343, exceeding the quote-first variants at 0.254 and 0.237. The analysis identifies evidence selection and answer construction as major factors affecting factual coverage.

\subsubsection{Answer-File Statistics}

The submitted generation file contains 8,003 answer objects across 119 narratives. Answer lengths range from 989 to 1,000 words, averaging 996.2. Each answer contains an average of 67.3 objects and 32.1 references, with reference counts ranging from 25 to 38. Citation verification left 1,948 objects without citations, representing 24.3\% of the total.

\subsection{Prompts used by the submitted systems}

The uploaded runs call language models with the templates below.
Text in braces is filled at runtime with the narrative, facets, passages, or claims. Experimental prompts for hybrid weight assignment, quote-first extraction, and claim rewriting were not used in the uploaded files.

\subsubsection{Sparse query planning}

Model: \texttt{gpt-4.1-mini}.

\begin{promptbox}
You are an information retrieval planner. Your job is to rewrite a complex topic into a sparse-friendly search query.

Given the topic below, produce JSON with:
- "subquestions": 3 to 6 short subquestions
- "entity_terms": important exact-match entities, dates, places, abbreviations
- "keywords": high-value lexical terms and near-synonyms
- "sparse_big_query": one single query string made by concatenating the subquestions and the best keywords
Rules:
- Keep exact names and dates unchanged.
- Prefer precise nouns and technical terms.
- Do not add unsupported facts.
- Keep the sparse_big_query under 120 words.
Topic:
{topic}
\end{promptbox}

\subsubsection{Hypothetical evidence sketches}

Model: \texttt{gpt-4.1-mini}.
The resulting passages are searched with BM25.

\begin{promptbox}
You generate dense-retrieval evidence sketches.

Given the topic below, output JSON with:
- "evidence_sketches": 2 to 4 short passages, each 40 to 80 words
Rules:
- Each sketch should describe one likely answer facet.
- Write in plain factual prose as if it were a relevant passage.
- It is okay if details are approximate; do not invent named entities that are not already implied by the topic.
- Cover different facets, not paraphrases of the same facet.
Topic:
{topic}
\end{promptbox}

\subsubsection{Sub-narrative generation}

The test run requests two independent decompositions and merges them.
The examples block contains two style examples from official RAG 2025 narratives.

\begin{promptbox}
You decompose first-person information needs into atomic sub-narratives for passage grading. Return only valid JSON.

Decompose the narrative below into atomic sub-narratives.
A sub-narrative is one specific, independently answerable facet of the user's need.
Follow the official TREC RAG style:
- Produce 6 to 10 sub-narratives (prefer 7-9 when the narrative is multi-faceted).
- Each "text" is short (typically under 15 words): a noun phrase or a focused question.
- Cover distinct facets; do not paraphrase the same facet.
- Stay faithful to the narrative; do not invent entities, events, or claims not implied by it.
- Mark importance as "vital" for core facets and "okay" for secondary ones.
- Roughly half or more should be "vital".
Return JSON with "sub_narratives", each containing "text" and "importance".
{examples}
Narrative:
{narrative}
\end{promptbox}

\subsubsection{Pointwise passage grade}

Model: \texttt{gpt-4.1-mini}.
The model returns a score from 0 to 100 for every passage in the batch.

\begin{promptbox}
You are a strict TREC passage relevance assessor.
Information need: {topic}
Important facets: {facets}
Score each passage independently from 0 to 100:
- 0-10: irrelevant or wrong intent
- 11-30: same broad topic but not useful for this information need
- 31-50: useful background/context only
- 51-70: directly answers part of one requested facet
- 71-85: strong, specific evidence for a requested facet
- 86-100: exceptionally direct, detailed, and useful evidence
Judge only what the passage states. Penalize keyword stuffing, unsupported claims, vague generic prose, tangents, and passages that merely repeat the question. Use the full scale and score every id.
Passages: {passages}
Return JSON only: {"grades":[{"id":0,"score":0}, ...]}
\end{promptbox}

\subsubsection{UMBRELA grade}

Models: Qwen3.5-9B and Ministral-3-14B.
The required output line is \texttt{\#\#final score: X}, where \(X\in\{0,1,2,3,4\}\).

\begin{promptbox}
Given a user first person narrative and a passage, you must provide a score on an integer scale of 0 to 4:
0 - the passage has nothing to do with the narrative,
1 - the passage seems related but contains no sub-narrative of an answer,
2 - the passage contains a detailed answer for 1 sub-narrative,
3 - the passage contains a detailed answer for 2 to 3 sub-narratives,
4 - the passage contains a detailed answer for 4 or more sub-narratives.
Narrative: {narrative}
Sub-narratives: {sub_narratives}
Passage: {passage}
A passage must cover a sub-narrative in a detailed, proper way to count; mere mentions do not qualify. Extra unrelated information lowers the score by 1, or by 2 when it is extensive. Give only the score as: ##final score: X
\end{promptbox}

\subsubsection{Listwise reranking}

Model: \texttt{gpt-5.6-sol}.
The model sees the top 30 passages, truncated to 1,800 characters each.

\begin{promptbox}
You are a strict TREC UMBRELA listwise reranker. Optimize sub-narrative coverage detail. Return valid JSON only.

Rank these passages for a TREC RAG retrieval task.
Narrative: {narrative}
Atomic sub-narratives (a passage is better if it answers MORE of these IN DETAIL; mere mentions do not count; lots of off-topic content should rank lower): {sub_narratives}
Passages: {passages}
Return ONLY valid JSON with every passage id exactly once, best first: {"ordered_ids":[0,1,2,...]}
\end{promptbox}

\subsubsection{Atomic claim extraction}

The first pass uses \texttt{gpt-5.6-sol}.
The dense second pass uses \texttt{gpt-4.1-mini} on overlapping evidence windows and under-covered sub-narratives.

\begin{promptbox}
Extract up to {n_ask} ATOMIC factual claims that help answer the information need. Each claim must be a single concrete fact (<= 22 words) that appears in the evidence (names, numbers, comparisons, mechanisms, causes, effects, policies, risks, actions, dates, quantities).
Do NOT summarize into vague themes. Prefer many specific facts over a few broad ones. Quote entities and numbers faithfully. Skip titles and meta text.
Every claim MUST cite at least one evidence id that contains the fact.
Return JSON: {"claims": [{"text": ..., "support": [evidence_ids]}]}.
Information need: {topic}
{aspects_or_priority_aspect}
Evidence: {numbered_evidence_sentences}
\end{promptbox}

\subsubsection{Claim-to-sentence conversion}

Each claim becomes one sentence, and each sentence may use only the citations attached to that claim.

\begin{promptbox}
Convert each claim into ONE short fluent sentence (12-28 words) that PRESERVES the specific fact (names, numbers, comparisons). Do not merge claims. Do not drop details. No titles or bracket tags in text.
Citations: use only allowed_citations for that claim (1-3 ints).
Information need: {topic}
Claims: {claims_with_allowed_citations}
Return JSON: {"sentences": [{"id": <claim id>, "text": ..., "citations": [ints]}]}.
\end{promptbox}

\subsubsection{Citation support and repair}

Model: \texttt{gpt-4.1-mini}.
The model sees up to 4,000 characters of the cited document.
Citations labeled \texttt{not\_support} are removed.
Citations labeled \texttt{partial\_support} are kept.

\begin{promptbox}
You are a strict TREC RAG support assessor. For each numbered sentence, decide whether the PASSAGE supports it. Reply with a JSON object {"labels": [...]} whose list has exactly one label per sentence, in order. Each label is "support" (passage fully supports the sentence), "partial_support" (passage supports part of it), or "not_support" (passage does not support it). No other text.

NARRATIVE: {narrative}
PASSAGE: {passage}
SENTENCES: {numbered_sentences}
Return {"labels": [...]} with exactly one label per sentence.
\end{promptbox}

\end{document}